\documentstyle[aps,twocolumn]{revtex}
\title{Almost-Hermitian Random Matrices:
Crossover from Wigner-Dyson to Ginibre Eigenvalue Statistics}
\author{
Yan V. Fyodorov$\S$\cite{leave}
Boris A. Khoruzhenko$\P$
and Hans-J\"{u}rgen Sommers$\S$       }
\address{
$\S$Fachbereich Physik, Universit\"at-GH Essen,
D-45117 Essen, Germany
        }
\address{
$\P$ School of Mathematical Sciences, Queen Mary \& Westfield
College, \\ University of London, London E1 4NS, U.K.
        }
\date{\today}
\begin{document}
\draft
\maketitle
\bigskip

\begin{abstract}
By using the method of orthogonal polynomials we analyze the statistical
properties of complex eigenvalues
of  random matrices
describing a crossover from Hermitian matrices
characterized by the Wigner-
Dyson statistics of real eigenvalues to strongly non-Hermitian ones
whose complex eigenvalues were studied by Ginibre.
 Two-point statistical measures
(as e.g. spectral form factor, number variance and
small distance behavior of the nearest neighbor
distance distribution $p(s)$) are studied in more detail.
In particular, we found that the latter function
may exhibit unusual behavior
$p(s)\propto s^{5/2}$ for some parameter values.
\end{abstract}
\pacs{PACS numbers: 05.45.+b}
\vfill

Complex eigenvalues of large random matrices
have recently attracted much interest
\cite{FS,QCD,Khor,Nels,FKS,pass,Oas,Edel,Efet,free}.
Initially, physicists used the complex eigenvalues
to describe the generic statistical properties  of
resonances in open quantum
chaotic systems \cite{Sok,FS} and characterize typical features of
dissipative chaotic quantum maps \cite{diss}.
 Complex and real asymmetric matrices
were also used to simulate the geometric and topological
properties of two-dimensional froth \cite{froth}
and to investigate chaotic dynamics
of asymmetric neural networks \cite{Som,Lehm1,net}.
Very recently
complex random matrices were shown to be relevant for
understanding of  spontaneous chiral symmetry breaking in quantum
chromodynamics \cite{QCD} and also emerged in
studies on advection of the passive scalar by a quenched random
velocity field\cite{pass}.

Traditional mathematical treatment of random matrices
with no symmetry conditions imposed
goes back to the pioneering work by Ginibre\cite{Gin}
who determined  all the correlation functions of the
eigenvalues
in an ensemble of complex matrices with independent Gaussian entries.
The progress in the field was rather slow but steady, see
\cite{Mehta,Gir,Som,Lehm1,diss,Forr,Edel,Oas}.

Surprisingly,  all these studies
completely disregarded the existence of a nontrivial regime
of {\it weak non-Hermiticity} recognized in our preceding works \cite{FS,FKS}.
 This regime occurs when the
imaginary part of typical eigenvalues is comparable
with the mean {\it separation} between
neighboring eigenvalues along the real axis.
 In particular, the
mean density
of complex eigenvalues derived in
\cite{FKS} turned out to be described by a
formula containing as two opposite limiting cases both the Wigner
semicircular density of real eigenvalues of the Hermitian matrices
and the uniform density of complex eigenvalues predicted
for usual non-Hermitian random matrices\cite{Gin,Mehta,Gir,Som,Edel}.
Very recently, Efetov \cite{Efet} showed the relevance of almost-Hermitian
random matrices to a very interesting problem of
 motion of flux lines in superconductors
with columnar defects\cite{Nels}. He also managed to derive
the density of complex eigenvalues for a related, but
different set of almost-symmetric real random
matrices. This development clearly shows that, apart from
being a rich and largely unexplored mathematical object,
almost-Hermitian random matrices enjoy direct physical applications
and deserve to be studied in more detail.

Although giving an important insight into the problem, both papers \cite{FKS}
and \cite{Efet} suffered from the same deficiency:
 the present state
of art in the application of the supersymmetry technique \cite{Efrev}
used in \cite{FKS,Efet}
gives little hope
of access to quantities describing {\it correlations}
between  different eigenvalues in the complex plane
 due to unsurmountable technical difficulties.
At the same time, it is just the existence of these
{\it universal} correlations
which makes the random matrix theory such a powerful and versatile
tool of research in different branches of modern theoretical physics,
see e.g. review \cite{Bohigas}.

In the present paper we
develop the rigorous mathematical theory
of almost-Hermitian Gaussian random matrices based on the method of
orthogonal polynomials. Such a method is free from the above mentioned
problem and allows us to study
correlation properties of the complex spectra to the same degree as
 is typical for earlier studied classes of random matrices.

To introduce the notion of almost-Hermitian random matrices,
let us recall that every complex matrix $\hat{J}$ can be decomposed
into a sum of its Hermitian and skew-Hermitian parts. Following this
we consider
an ensemble of random $N\times N$ complex matrices
$\hat{J}=\hat{H}+iv\hat{A}$,
where $\hat H $ and
$\hat A$ are taken {\it independently} from a
Gaussian Unitary Ensemble (GUE) of {\it Hermitian } matrices
with the probability density
${\cal P}(\hat{X})=Q_N^{-1}
\exp{\Big(-N/2J_0^{2}\ \mbox{Tr}\ \hat{X}^2 \Big)}$,
$\hat X = \hat X^\dagger$.
The parameter $v$, $0\le v^2\le 1$, is a natural measure of
non-Hermiticity in our ensemble. When $v^2=0$, $\hat J$ belongs to the GUE
and when $v^2=1$, $\hat{J}$ belongs to the ensemble of complex matrices
considered by Ginibre\cite{Gin}. If $v^2$ is fixed, as
$N\to \infty$ the eigenvalues of $\hat{J}$ are uniformly distributed
in an ellipse in the complex plane\cite{Gir,Som} with the imaginary part being
typically of the order of $J_0$. This is just the
 regime of strong non-Hermiticity.
 In contrast, the regime of weak non-Hermiticity
is determined by the condition $v^2=O(1/N)$ when $N\to \infty$, see
the more detailed discussion in
\cite{FKS}.  In other words,
we scale the parameter $v$ with the matrix dimension $N$
as $2v=\alpha /\sqrt N$ and consider $\alpha$ to be fixed
as $N \to \infty$.

Let us now introduce a new parameter $\tau=(1-v^2)/(1+v^2)$
and choose the scale constant $J_0^2$ to be equal to
$(1+\tau)/2$, for the sake of convenience.
The parameter $\tau$ controls the magnitude of
correlation between
$J_{jk}$ and $J_{kj}^*$: $\langle J_{jk}J_{kj}^* \rangle = \tau /N $,
hence
the degree of non-Hermiticity.
This is easily seen from the probability density function
for our ensemble of the random matrices $\hat{J}$:
\begin{eqnarray}
{\cal P}(\hat{J})
=C_N^{-1} \exp \Big[-\frac{N}{(1-\tau^2)} \ \mbox{Tr}
(\hat{J}\hat{J}^\dagger -\tau \ \mbox{Re}\  \hat{J}^2) \Big],
\label{P(J)}
\end{eqnarray}
where $C_N=[\pi^2(1-\tau^2)/N^2]^{N^2/2}$.
All the $J_{jk}$  have zero mean and variance
$\langle |J_{jk}|^2 \rangle =1/N$ and only
$J_{jk}$ and $J_{kj}$ are pairwise correlated.
If $\tau =0$ all the $J_{jk}$
are mutually independent and
we have maximum non-Hermiticity.
When $\tau $ approaches unity, $J_{jk}$ and
$J_{kj}^*$ are related via $J_{jk}=J_{kj}^*$
and we are back to an ensemble of Hermitian matrices.

Our first goal is to determine the $n$-eigenvalue correlation functions
in the ensemble of random matrices specified by Eq.\ (\ref{P(J)}).
The density of the joint distribution of
eigenvalues in the ensemble is given by
\begin{eqnarray}\label{P(Z)}
\lefteqn{
{\cal P}_N(Z_1, \ldots, Z_N)= \frac{N^{N(N+1)/2}}{\pi^N 1!
\cdots N! (1-\tau^2)^{N/2}}
\times        }  \\ \nonumber
 & &
\exp \Big\{\frac{-N}{1-\tau^2} \sum_{j=1}^N
\Big[|Z_j|^2 - \frac{\tau}{2}(Z_j^2+{Z_j^*}^2) \Big]  \Big\} \
\prod_{j<k}|Z_j-Z_k|^2 .
\end{eqnarray}
One can derive Eq.~(\ref{P(Z)}) following more or less standard
reasoning.
By definition, one obtains ${\cal P}_N(Z_1, \ldots Z_N) $ integrating
${\cal P}(\hat{J})$  over the surface of all complex matrices
whose eigenvalues are $Z_1, \ldots Z_N$. We disregard the matrices
with multiple eigenvalues because the set of these matrices
has zero probability measure of Eq.~(\ref{P(Z)}).
Every complex matrix with distinct eigenvalues $Z_1, \ldots , Z_N$
can be decomposed as
$ \hat{J}=\hat{U}(\hat{Z}+\hat{R})\hat{U}^{\dagger}$,
where $\hat{Z}=\mbox{diag}\{ Z_1, \ldots Z_N \}$,
$\hat {U}$ is  a unitary matrix, and
$\hat{R}$ is a strictly upper-triangular (with zero diagonal)
complex matrix.
In this decomposition $\hat{Z}$ is determined modulo
permutation of the eigenvalues and $\hat{U}$ is determined
modulo multiplication to the right by a diagonal unitary matrix.
Therefore if we label the eigenvalues and
require the first
non-zero element in each column of $\hat{U}$ to be positive, then
the decomposition is unique. We can parameterize the above mentioned
surface by the matrices $\hat R$ and $\hat U$.
The Jacobian of the corresponding transformation is given by
the squared modulus of the Vandermonde determinant [the
last factor on the r.h.s. in Eq.~(\ref{P(Z)})]. The integrand
${\cal P}(\hat J)$ does not depend on $\hat U$ being a Gaussian
function in $\hat R$.
So, integrating out these two variables is straightforward.
The resulting expression (with the requirement of labeling
the eigenvalues being removed) is Eq.~(\ref{P(Z)}).

The form of the distribution
Eq.~(\ref{P(Z)}) allows one to employ
the powerful method of orthogonal polynomials \cite{Mehta}.
Let $H_n(z)$ denote the
$n$-th Hermite polynomial,
\begin{eqnarray} \label{H}
H_n(z)=\frac{(\pm i)^n}
{\sqrt{2\pi}}\! \exp{\left(\! \frac{z^2}{2}\! \right)}
\int_{-\infty}^{\infty}\!
dt\  t^n \exp{\left(-\frac{t^2}{2}\mp izt\right)}.
\end{eqnarray}
The
crucial observation  borrowed from the paper
 \cite{orth} is that the polynomials
\begin{eqnarray}\label{p_n}
p_n(Z)=\frac{\tau^{n/2} \sqrt{N} }
{\sqrt{\pi }\sqrt{ n!}(1-\tau^2 )^{1/4}}
H_n\left( \sqrt{ \frac{N}{\tau}}Z\right),
\end{eqnarray}
$n=0,1,2, \ldots $,
are orthogonal in the {\it complex plane}
$Z=X+iY$
with the weight function
\[
w^2(Z)=\exp{\left\{-\frac{ N}{(1-\tau^2)}\left[|Z|^2 -
\frac{\tau}{2}(Z^2+{Z^*}^2) \right]\right\}},
\]
i.~e. $
\int d^2Z p_n(Z)p_m(Z^*)w^2(Z)  = \delta_{nm}$, where
$d^2 Z = dX dY$.
The standard machinery of the method of orthogonal polynomials\cite{Mehta}
yields the $n$-eigenvalue correlation functions
\begin{eqnarray}
R_n(Z_1,...,Z_n)=\frac{N!}{(N-n)!}\int d^2Z_{n+1}...d^2Z_N
{\cal P}_N\{Z\}
\label{R_n}
\end{eqnarray}
in the form
\begin{eqnarray*}
R_n(Z_1,...,Z_n)&=&\det \left[ K_N(Z_j,Z_k^*)\right]_{j,k=1}^n,
\end{eqnarray*}
where the kernel $K_N(Z_1,Z_2^*)$ is given by
\begin{eqnarray}
K_N(Z_1,Z_2^*)&=w(Z_1)w(Z_2^*)&\sum_{n=0}^{N-1}p_n(Z_1)p_n(Z_2^*).
\label{K}
\end{eqnarray}

With Eqs.\ (\ref{p_n})--(\ref{K}) in hand, let us first examine
the regime of strong non-Hermiticity, i.e. the case when
$\lim_{N\to \infty} (1-\tau ) > 0$. In this regime the following
Mehler's formula wich follows from Eq.\ (\ref{H}) is helpful:
\begin{eqnarray*}
\lefteqn{
         \sum_{n=0}^{\infty}
         \frac{\tau^n}{n!} H_n(z_1)H_n(z_2)=
       } \\
 & &
\frac{1}{   \sqrt{  1-\tau^2 }   }
\exp{
     \left\{
            \frac{\tau}{1-\tau^2}
     \left[
            z_1z_2 -\frac{\tau}{2}
     \left(
            z_1^2+z_2^2
     \right)
     \right]
     \right\}
    }.
\end{eqnarray*}
Exploiting this relation one easily shows that
in the regime of strong non-Hermiticity all the
correlation functions $R_n(Z_1,Z_2, \ldots, Z_n)$ coincide,
after trivial rescaling $Z_j \to  (1-\tau^2)^{-1/2}Z_j$,
with those found by Ginibre\cite{Gin}.

Now we move on to the regime of weak non-Hermiticity:
We will
demonstrate that in this regime
new non-trivial correlations occur
on the scale:
$\mbox{Im} Z_{1,2}=O(1/N)$, $\mbox{Re} Z_1-\mbox{Re}
 Z_2=O(1/N)$.
Correspondingly, we
introduce new variables $x,y_1,y_2,\omega$ in such a way that:
$x=\mbox{Re}\left(Z_1+Z_2)\right/\! 2$,
$y_{1,2}=N\mbox{Im}\left(Z_{1,2} \right)$,
$\omega=N\mbox{Re}\left(Z_1-Z_2\right)$,
 and consider them finite when performing the limit $N\to
\infty$.

Substituting Eq.(\ref{H}) into Eq.(\ref{K}) and
using the above definitions
we can explicitly perform the limit $N\to \infty$,
taking into account that $\lim_{N\to \infty} N(1-\tau)= \alpha^2/2$.
The details of the procedure will be given elsewhere\cite{unpub}.
As a result, the kernel $K_N(Z_1,Z_2^*)$
 is given by the following expression:
\begin{eqnarray}\label{kern}
K_N(Z_1,Z_2^*)=g_{\alpha}(y-i\omega/2)\times \\ \nonumber\frac{N}{\pi\alpha}
\exp{\left\{- \frac{y_1^2+y_2^2}{\alpha^2}+\frac{ix (y_1-y_2)}{2}
\right\}},
\end{eqnarray}
where  we have intoduced introduced the notation $ y=(y_1+y_2)/2$ and
\begin{equation}\label{g}
g_{\alpha}(y)=\int_{-\pi\nu_{sc}(x)}^{\pi\nu_{sc}(x)}\frac{du}{\sqrt{2\pi}}
\exp{\left[-\frac{\alpha^2u^2}{2}-2uy\right]},
\end{equation}
with
$\nu_{sc}(x)=\frac{1}{2\pi}\sqrt{4-x^2}$ standing for the Wigner
semicircular density of
real eigenvalues of the Hermitian part $\hat{H}$ of the matrices $\hat{J}$.

Equation (\ref{kern})
 constitutes the most important result of the present publication.
The kernel $K_N$ given by Eq.\ (\ref{kern})
determines all the properties of complex eigenvalues in the regime of
weak non-Hermiticity. For instance,
the mean value of the density
$\rho(Z)= \sum_{i=1}^N\delta^{(2)}(Z-Z_i)$
of complex eigenvalues $Z=X+iY$ is  given by $
\langle\rho(Z)\rangle= K_N(Z,Z^*)$
Putting $y_1=y_2$ and $\omega=0$ in Eqs.\ (\ref{kern})--(\ref{g})
we immediately recover the density of
complex eigenvalues of almost-Hermitian random matrices obtained for
the first time in our preceding publication \cite{FKS} and reproduced
recently in \cite{Efet}.

One of the most informative statistical measures of the spectral correlations
is the `connected' part of the two-point correlation function of eigenvalue  
densities:
\begin{eqnarray}
\left\langle \rho(Z_1)\rho(Z_2)\right\rangle_c
=\left\langle
\rho(Z_1)\right\rangle \delta^{(2)}(Z_1-Z_2)-Y_2(Z_1,Z_2),
\end{eqnarray}
with the so-called {\it cluster function} $Y_2(Z_1,Z_2)$
being equal to $Y_2(Z_1,Z_2)=\left|K_N(Z_1,Z^*_2)\right|^2$.
In particular, it determines the variance $\Sigma^2(D)=
\langle n(D)^2\rangle-\langle n(D)\rangle^2$ of the number
$n=\int_D d^2Z\rho(Z)$
of complex eigenvalues in any domain $D$ in the complex plane.
 It also
allows one to find the small-distance behaviour
of the so-called nearest-neighbour distance distribution :
  $p(z_0,s)$ , i.e. the probability
density to have an eigenvalue at a point $z_0$ of the complex plane
such that its closest neighbour is at the point $z_1$
separated by the distance
$|z_1-z_0|=s/N$, see \cite{Mehta,diss,Oas}.

We see, that in the regime of weak non-Hermiticity
the cluster function depends on the variable
$x=\mbox{Re}(Z_1+Z_2)/2$ only via the semicircular density $\nu_{sc}(x)$.
The parameter
$a=2\pi\nu_{sc}\alpha$ controls the deviation from Hermiticity.
When $a\to 0$ the cluster function tends to GUE form
$Y_2(y_1,y_2,\omega)=N^2\delta(y_1)
\delta(y_2)\sin^2{(\pi\nu_{sc}(x)\omega)}/(\pi\omega)^2$,
whereas in the opposite case $a\gg 1$ the limits of integration in Eq.(\ref{g})
can be effectively put to $\pm\infty$, and the corresponding Gaussian
integration yields the Ginibre \cite{Gin} result:
$Y_2(Z_1,Z_2)=(N/\pi \alpha)^{2}
\exp\{-N^2|Z_1-Z_2|^2/{\alpha}^2\}$.

 Fourier-transforming the  cluster function
over all arguments $y_1,y_2$ and $\omega$ we find
the following expression for the {\it spectral form-factor}:
\begin{eqnarray}\label{for}
b(q_1,q_2,k)=
N^2\exp\{-\frac{\alpha^2}{8}\left(q_1^2+q_2^2+2k^2\right)\}
\times  \\ \nonumber
\frac{\sin{\left[\frac{ \pi\alpha^2
(q_1+q_2)}{2}\left(\nu_{sc}(x)-\frac{|k|}{2\pi}
\right)\right]}}
{\pi \alpha^2(q_1+q_2)/2}\theta(2\pi\nu_{sc}(x)-|k|)\,
\end{eqnarray}
where $\theta(u)=1$  for $u>0$ and zero otherwise.

As we already mentioned, the knowledge of the cluster function
 allows one to determine the variance of the number of
eigenvalues in any domain $D$ of the complex plane. In the general case
this expression is not very transparent, however. For this reason we
present the explicit form for the simplest case,
choosing  the domain $D$ to be the infinite strip of width $L$
(in units of mean spacing along the real axis $\Delta=(\nu_{sc}(x)N)^{-1}$)
oriented perpendicular to the real axis:
$0<\mbox{Re}Z<~L\Delta;\quad -\infty
<\mbox{Im}Z<\infty$. One finds:
\begin{equation}
\label{var}
\Sigma_2(L)=
L-\frac{2}{\pi^2}\! \int_0^1\! \frac{dk}{k^2}(1-k)\sin^2{(\pi k L)}
e^{-(\frac{a\pi k}{2})^2 }
\end{equation}

As long as $a\ll L$ the function $\Sigma_2(L)$ slightly
deviates from the corresponding expression for the Hermitian matrices
\cite{Mehta,Bohigas}.
Let us briefly discuss the gross  features of the number variance
 at $a\gg 1$ when we expect essential differences from the Hermitian case.
Normally one is interested in large $L$ asymptotics: $L\gg 1$.
In that case the upper limit of the integral in Eq.(\ref{var})
can be set to infinity. This implies that in a large domain $1\ll L\sim a$
the number variance grows linearly like
 $\Sigma(L)=Lf(2L/a)$ with
$$
f(u)=1+\frac{2}{\sqrt{\pi}}\left\{\frac{1}{2u}\left(1-e^{-u^2}\right)-
\int_{0}^{u}dte^{-t^2}\right\}.
$$
For $u=2L/a\ll 1$ we have simply $f\approx 1$. For $u\gg 1$ we have
$f~\approx~(\sqrt{\pi}u)^{-1}$. Thus, $\Sigma_2(L)$ slows down:
$\Sigma_2(L)\approx a/2\sqrt{\pi}$. Only for exponentially large $L$ such that
$\ln{(L/a)}\gtrsim a$ does
the term proportional to $k^{-1}$ in the integrand of
Eq.(\ref{var}) contribute significantly. Then it leads
to the slower logarithmic
increase:
$$
\Sigma_2(L)= \frac{a}{2\sqrt{\pi}}+\frac{1}{\pi^2}
\left(\ln{\left(\frac{2L}{\pi a}\right)}-\frac{\gamma}{2}\right)
+ \mbox{corrections},
$$
where $\gamma$ is Euler's constant. This behaviour is reminiscent of that
typical for real eigenvalues of the Hermitian matrices.

One can also derive the following expression for small-distance
 behaviour of the  probability density to have one
eigenvalue at the point $z_0=X+iy_0/N$ and its
closest neighbor at the distance $|z_1-z_0|=s/N$:
\begin{eqnarray} \label{nns}
p_{\alpha}(z_0,s\ll 1)=\frac{g^2_{\alpha}(y_0)\frac{\partial^2}{\partial y_0^2}
\ln{g_{\alpha}(y_0)}}{(2\pi\alpha)^2}e^{-2\frac{y_0^2}{\alpha^2}}\times \\
\nonumber 2s^3 \int_0^{\pi}d\theta
\exp{\left[-\frac{2}{\alpha^2}(y_0-s\cos{\theta})^2\right]}
\end{eqnarray}.

It is easy to see that performing the limit $\alpha\to 0 $ we are back to
familiar GUE quadratic level repulsion: $p_{\alpha\to 0}(z_0,s\ll 1)
\propto \delta(y_0) s^2$, whereas in the opposite limit $\alpha\to \infty$
one has: $p_{\alpha\to \infty}(z_0,s\ll 1)
=\frac{2}{\pi} (s/\alpha^2)^3$ in agreement with the cubic repulsion generic  
for strongly non-Hermitian random matrices\cite{Gin,diss,Oas}. In general,
the expression in Eq.(\ref{nns})
describes a smooth crossover between the two regimes.

To this end, an interesting regime occurs for the
`observation points' $z_0$ situated sufficiently far
from the real axis: $|y_0|\gg 2^{-3/2}\alpha$.
It turns out that for such points the distribution
$p(z_0,s)$ displays an unusual $s^{5/2}$ power-law behaviour:
\begin{equation}
p_{\alpha}(z_0,s\ll 1)=\frac{g^2_{\alpha}(y_0)\frac{\partial^2}{\partial y_0^2}
\ln{g_{\alpha}(y_0)}}{2(2\pi)^{3/2}\alpha |y_0|^{1/2}}
e^{-4\frac{y_0^2}{\alpha^2}} s^{5/2}
\end{equation}
in the parametrically large domain of distances:
$\alpha^2/4|y_0|\ll s\ll 2|y_0|$. Unfortunately,
the low density of complex eigenvalues in
such points (reflected by the presence of the
Gaussian factor in the expression above) would make any
direct verification of such an unconventional
behaviour to be a difficult numerical problem.

Finally, we would like to make the following remark.
So far all the expressions were derived rigorously for the Gaussian
almost-Hermitian random matrices by using the probability
measure of Eq.(\ref{P(J)}) as a starting point.
However, employing the
supersymmetry technique one is able to give strong,
albeit not mathematically rigorous arguments, \cite{unpub}
that all the results obtained in this way are fairly
universal and equally applicable for a much broader class of
almost-Hermitian matrices $\hat{J}=\hat{H}+iv\hat{A}$,
provided the semicircular density $\nu_{sc}(x)$ is replaced by
the actual eigenvalue density $\nu(x)$ of the Hermitian matrix $\hat{H}$
on the real line at point $x$.

In conclusion, we have developed a rigorous mathematical theory
of almost-Hermitian Gaussian random matrices based on the method of
orthogonal polynomials. This has enabled us to study in great detail
the two-point
correlations between complex eigenvalues in the novel crossover
regime between Hermitian and strongly non-Hermitian random matrices.

 The financial support
       by SFB-237(Y.V.F and H.-J.S.) and EPRSC Research Grant GR/L31913
       (Y.V.F. and B.A.K.) is acknowledged with thanks.
 Y.V.F. is grateful to the School of Mathematical Sciences, Queen Mary\&
Westfield
College, University of London for the warm hospitality extended to him during
his visit when a part of this work was done. B.A.K would like to thank
Werner Kirsch for
stimulating his interest in non-Hermitian
random matrices.

\end{document}